\documentclass[prl,a4paper,twocolumn,groupedaddress,english]{revtex4-1}
\usepackage{babel}
\usepackage{xcolor}
\usepackage{graphicx}
\usepackage{amsmath}
\usepackage{amssymb}
\usepackage{wasysym}
\usepackage{hhline}

\newcommand{\bc}{\begin{center}}
\newcommand{\ec}{\end{center}}
\newcommand{\be}{\begin{equation}}
\newcommand{\ee}{\end{equation}}
\newcommand{\bea}{\begin{eqnarray}}
\newcommand{\eea}{\end{eqnarray}}
\newcommand{\dagga}{{\phantom{\dagger}}}

\definecolor{darkblue}{rgb}{0.1,0.2,0.6}
\definecolor{darkred}{rgb}{0.8,0.1,0.2}
\usepackage[colorlinks,citecolor=darkblue,linkcolor=darkred,urlcolor=darkblue]
{hyperref}
\usepackage[all]{hypcap} 

\makeatletter
\adddialect\l@English\l@english
\makeatother

\newcommand{\ket}[1]{|\hskip -0.15cm\,#1\rangle}

\newcommand{\co}{(Color online) }

\definecolor{commentcolor_nl}{rgb}{0.1,0.2,0.6}
\definecolor{commentcolor_md}{rgb}{1,0,0}
\definecolor{commentcolor_sc}{rgb}{0,0,1}
\definecolor{commentcolorD}{rgb}{1,0.1,.1}
\definecolor{todocolor}{rgb}{0.8,0.1,0.2}

\begin{document}
\title{Disorder-induced Revival of the Bose-Einstein Condensation\\
in Ni(Cl$_{1-x}$Br$_x$)$_2$-4SC(NH$_2$)$_2$ at High Magnetic Fields}
\author{Maxime Dupont}
\author{Sylvain Capponi}
\affiliation{Laboratoire de Physique Th\'eorique, IRSAMC, Universit\'e de
Toulouse,
{CNRS, 31062 Toulouse, France}}
\author{Nicolas Laflorencie}
\affiliation{Laboratoire de Physique Th\'eorique, IRSAMC, Universit\'e de
Toulouse,
{CNRS, 31062 Toulouse, France}}

\begin{abstract}
Building on recent NMR experiments [A. Orlova \textit{et al.}, \href{http://journals.aps.org/prl/abstract/10.1103/PhysRevLett.118.067203}{Phys. Rev. Lett. {\bf 118}, 067203 (2017)}], we theoretically investigate the high magnetic field regime of the disordered quasi-one-dimensional $S=1$ antiferromagnetic material Ni(Cl$_{1-x}$Br$_x$)$_2$-4SC(NH$_2$)$_2$. The interplay between disorder, chemically controlled by Br-doping, interactions, and the external magnetic field,
leads to a very rich phase diagram.  Beyond the well-known antiferromagnetically ordered regime, analog of a Bose condensate of magnons, which disappears when $H\ge 12.3$ T, we unveil a resurgence of phase coherence at higher field $H\sim 13.6$ T, induced by the doping.
Interchain couplings stabilize finite temperature long-range order whose extension in the field -- temperature space is governed by the concentration of impurities $x$. Such a ``mini-condensation'' contrasts with previously reported Bose-glass physics in the same regime, and should be accessible to experiments.
\end{abstract}
\maketitle
{\it Introduction.---}
Interacting quantum systems in the presence of disorder have been intensively studied for several decades, leading to fascinating physics, {\it{e.g.}} the Kondo effect~\cite{hewson_kondo_1993}, the many-body localization transition~\cite{nandkishore_many-body_2015}, or the superfluid to Bose-glass (BG)~\cite{giamarchi_localization_1987,fisher_boson_1989} transition at finite disorder for lattice bosons~\cite{alvarez_zuniga_critical_2015,ng_quantum_2015,gurarie_phase_2009}.
While counterintuitive, in some situations disorder may enhance long-range order, as discussed for inhomogeneous superconductors~\cite{taguchi_increase_2006,tsai_optimal_2008,burmistrov_enhancement_2012}. Perhaps even more surprisingly, doping gapped antiferromagnets with a finite concentration of magnetic or non-magnetic impurities can fill up the bare spin gap with localized levels~\cite{xu_holes_2003,kenzelmann_structure_2003,schmidiger_emergent_2016} which may eventually order, in the strict sense of macroscopic long-range order (LRO) at low temperature. Such an impurity-induced ordering mechanism of the type ``order from disorder''~\cite{villain_order_1980,shender_dilution-induced_1991} has been experimentally observed for a large number of spin-gapped compounds~\cite{bobroff_impurity-induced_2009}: weakly coupled $d=1$ systems such as  spin-Peierls chains CuGeO$_3$~\cite{hase_effects_1993,manabe_antiferromagnetic_1998}, spin ladders SrCu$_2$O$_3$~\cite{azuma_switching_1997} and BiCu$_2$PO$_6$~\cite{bobroff_impurity-induced_2009}, Haldane chains PbNi$_2$V$_2$O$_8$~\cite{uchiyama_spin-vacancy-induced_1999}, as well as weakly coupled dimers in TlCuCl$_3$~\cite{oosawa_impurity-induced_2002}. Nevertheless, only a few studies have focused on the effect of
an external field~\cite{mikeska_field-induced_2004,nohadani_bose-glass_2005,roscilde_quantum_2005,roscilde_field-induced_2006,yu_magnetic_2010,casola_field-induced_2013}.

\begin{figure}
    \includegraphics[width=.85\columnwidth,clip,angle=0]{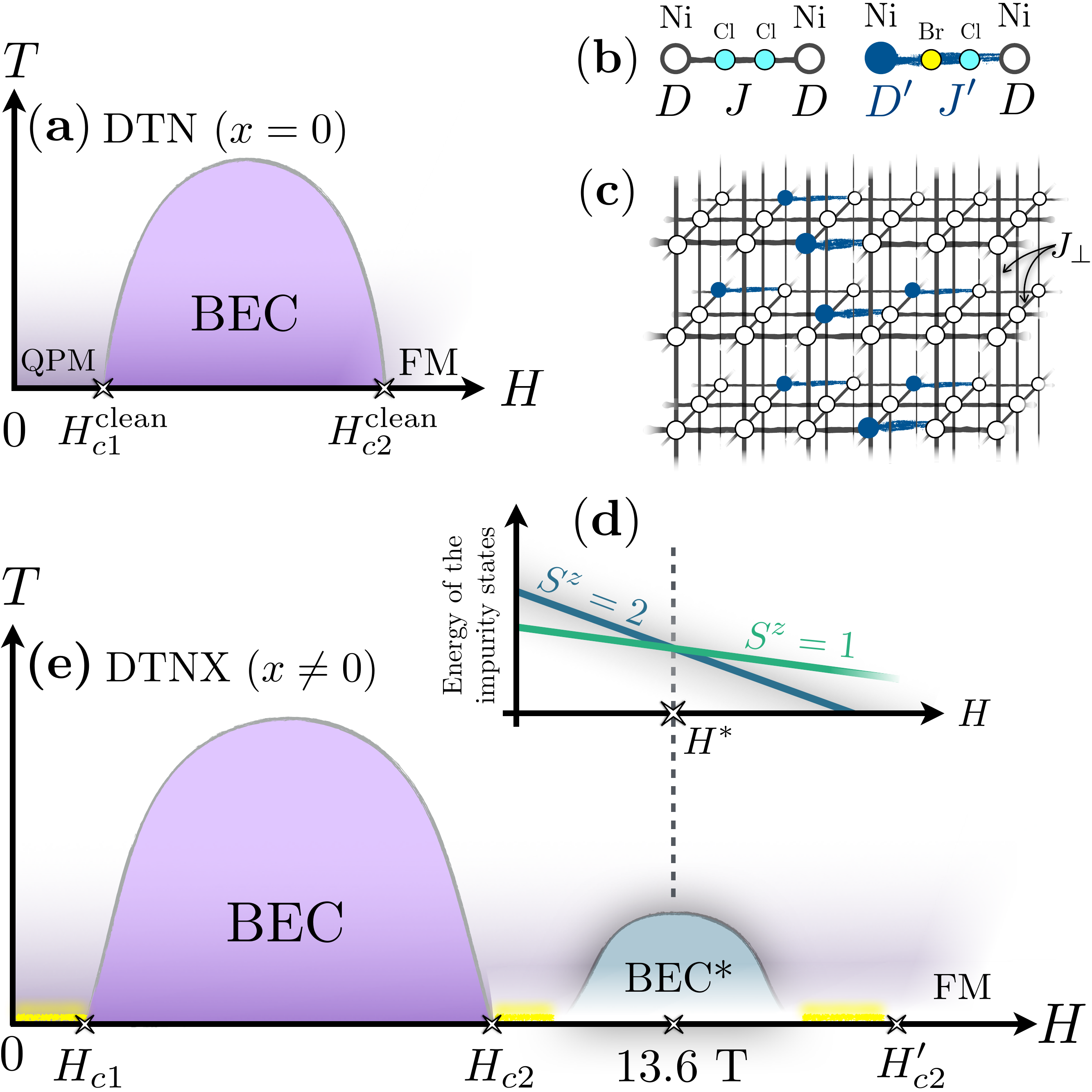}
    \vskip -.25cm
    \caption{\co~(a) Schematic temperature $T$ - magnetic field $H$ phase diagram of the clean DTN compound,  showing the BEC dome surrounded by a quantum paramagnet (QPM) and a polarized ferromagnet (FM). (b-c) The two types (clean and Br-doped) of $S=1$ dimers and their arrangement in a three-dimensional array of coupled chains modelling DTNX,
    see Eq.~\eqref{eq:DTNX}. (d) Field-induced energy level crossing of a Br-doped $S=1$ dimer. (e) Schematic $T$ -  $H$  phase diagram for DTNX at low doping, with an impurity-induced BEC$^*$ dome revival in the vicinity of the crossover field $H^*$. Yellow baselines show the regions where a BG is expected.}
     \label{fig:sketch}
\end{figure}

In this Letter, building on recent nuclear magnetic resonance (NMR) experiments~\cite{orlova_nuclear_2016}, we achieve a realistic theoretical study of the high magnetic field regime of Ni(Cl$_{1-x}$Br$_x$)$_2$-4SC(NH$_2$)$_2$ (DTNX): a three-dimensional antiferromagnetic (AF) system made of weakly coupled chains of $S=1$ spins subject to single-ion anisotropy [panels (b-c) of Fig.~\ref{fig:sketch}]. Note that the $S=1$ chains are not of Haldane type, due to the large anisotropy $D$~\cite{wierschem_characterizing_2014}. In the absence of chemical disorder ($x=0$), NiCl$_2$-4SC(NH$_2$)$_2$ (DTN) provides a very good realization of magnetic field-induced Bose-Einstein condensation (BEC) in a quantum spin system~\cite{affleck_bose_1991,giamarchi_coupled_1999,giamarchi_boseeinstein_2008,zapf_bose-einstein_2014} between two critical field $H_{c1}^{\rm clean}=2.1$ T and $H_{c2}^{\rm clean}=12.32$~T~\cite{paduan-filho_field-induced_2004,zapf_bose-einstein_2006,blinder_nuclear_2016}, see Fig.~\ref{fig:sketch} (a). Disorder induced by Br-doping locally changes the amplitude of the AF interaction between nearest neigbors Ni ($S=1$) atoms, at random positions along the chains. This is expected to bring new physics above $H_{c2}$ where a Bose-glass (BG) regime~\cite{giamarchi_localization_1987,fisher_boson_1989} with a disordered many-body groundstate
was recently reported~\cite{yu_bose_2012,yu_quantum_2012}. However, the Br-doped bonds introduce a new energy scale in DTNX, as shown by an enhanced NMR relaxation around a cross-over field $H^*\simeq 13.6$~T~\cite{orlova_nuclear_2016}. In close analogy with impurity-induced LRO at zero field~\cite{bobroff_impurity-induced_2009},  we show using large scale quantum Monte Carlo (QMC) simulations that in the vicinity of $H^*$ localized states hosted by doped bonds can interact and display macroscopic coherence,  as sketched in Fig.~\ref{fig:sketch}.

{\it Theoretical modelling of DTNX.---} Recent neutron~\cite{povarov_dynamics_2015} and NMR~\cite{orlova_nuclear_2016} experiments on DTNX at various Br concentration $0.04\le x \le 0.13$ have both shown the existence of a localized level above $H_{c2}$. Building on NMR data~\cite{orlova_nuclear_2016}, the microscopic parameters for Br-doped bonds [there are two non-equivalent Cl sites in each J bond, but only one of these can be doped by a Br, see panel (b) of Fig.~\ref{fig:sketch}] can be precisely determined in order to match the observed  spin relaxation peak at $H^*\simeq 13.6$ T, attributed to the crossing between $S^z=2$ and $S^z=1$ levels of impurity states [panel (d) of Fig.~\ref{fig:sketch}], combined with the local magnetizations from NMR shifts. DTNX is therefore described by the following $S=1$ model~\cite{orlova_nuclear_2016,dupont_long_2016}:
\bea
{\cal{H}}&=&\sum_i\Bigl[\sum_nJ_{i,n}{\boldsymbol{S}}_{i,n}\cdot {\boldsymbol{S}}_{i+1,n} +J_\perp\sum_{\langle n\,m\rangle}{\boldsymbol{S}}_{i,n}\cdot {\boldsymbol{S}}_{i,m}\nonumber\\
&+&\sum_n D_{i,n}\left(S_{i,n}^{z}\right)^2-g\mu_B H S_{i,n}^z\Bigr],
\label{eq:DTNX}
\eea
where the various parameters are shown in Fig.~\ref{fig:sketch} (b-c). Along the chain direction, undoped bonds display an AF exchange $J_{i,n}=J=2.2$ K while for Br-doped bonds (in concentration $2x$) $J_{i,n}=J'=5.32$ K. Single-ion anisotropies are $D_{i,n}=D=8.9$~K for clean sites and $D_{i,n}=D'=3.2$~K for the sites adjacent to a doped Br atom, here on the left side of the doped bond, see Fig.~\ref{fig:sketch} (b). Since the transverse bonds which couple the chains in a three-dimensional ($3d$) array are not directly affected by Br-doping,  interchain coupling between nearest-neigbor sites $\langle n\,m\rangle$ is assumed to take its clean value $J_\perp=0.18$ K. In the following, we use $g=2.31$ for the gyromagnetic factor, such that the clean upper critical field $H_{c2}^\mathrm{clean}=(D+4J+8J_\perp)/(g\mu_B)=12.32$ T~\cite{blinder_nuclear_2016}.

The coupling energy of a doped $S=1$ dimer being larger than for the undoped case $(J'/J=2.42)$, we first analyze an isolated ``impurity dimer'' [right-hand side of panel (b) in Fig.~\ref{fig:sketch}] embedded in a clean system.  Starting at high field, upon decreasing $H$ the polarized state ($S^z=2$) ${\ket{\uparrow\uparrow}}$ crosses the $S^z=1$ state at $H^*$ [panel (d) in Fig.~\ref{fig:sketch}]. Contrary to a clean system where the $S^z=1$ state would disperse, here its dynamics is described by a tight-binding model with a boundary impurity potential well~\cite{dupont_long_2016} of depth $\Delta_{\rm imp}=J'-J+\frac{D'-D}{2}\left(\sqrt{1+\left(\frac{2J'}{D-D'}\right)^2}-1\right)\simeq 6.3$~K, which localizes the $S^z=1$ state. The energy of such a bound state can be computed analytically in the limit of small interchain coupling $J_\perp\ll J$ and large impurity potential $\Delta_{\rm imp}\gg J_\perp$, thus yielding~
\be
H^*\approx D+\Delta_{\rm imp}+2J+4J_\perp+\frac{J^2}{\Delta_{\rm imp}}+\frac{4J_\perp^2}{\Delta_{\rm imp}}\simeq 13.6 ~{\rm T},
\ee
~perfectly matching the experiments~\cite{orlova_nuclear_2016}.

For a small but finite concentration $x$ of Br ions, around the crossover field $H^*$ we are left with a collection of localized states which are randomly placed in the $3d$ system of coupled chains. Using the above parameters, the localization length was determined to be very short~\cite{orlova_nuclear_2016}, in units of lattice spacings $\xi_{\parallel}\simeq 0.48$ along the chain and
$\xi_{\perp}\simeq 0.17$ in the transverse directions. Despite its random distribution in real space, this set of localized two-level systems is expected to experience an effective unfrustrated pair-wise coupling, exponentially suppressed with the distance~\cite{sigrist_low-temperature_1996,imada_scaling_1997,yasuda_site-dilution-induced_2001,laflorencie_doped_2003,doretto_quantum_2009,lavarelo_magnetic_2013},
and their density is controlled by a chemical potential, proportional to the external field $\mu=g\mu_B(H-H^*)$. From such considerations, a minimal toy-model with hard-core bosons (HCB) would read:
\be
{\cal H}_{\rm toy}=\sum_{\langle i j\rangle} t_{ij}\left(b_{i}^{\dag}b_{j}^{\dagga}+{{\rm h.c.}}\right)-\mu\sum_i b_{i}^{\dag}b_{i}^{\dagga},
\label{eq:eff}
\ee
where non-frustrated hopping terms $t_{ij}$ between neighbors are built from the effective pair-wise mechanism derived in Ref.~\cite{orlova_nuclear_2016}, and
for which one might expect a global phase coherence at low enough temperature~\cite{laflorencie_random-exchange_2006,bobroff_impurity-induced_2009}.

In the following, we  investigate in details such an ``order from disorder'' mechanism using large scale QMC simulations first for the realistic microscopic $S=1$ model, see Eq.~\eqref{eq:DTNX}, and then compare it with the above toy-model description  Eq.~\eqref{eq:eff}.\\

{\it Impurity-induced LRO at $H^*=13.6$ T.---} Using QMC Stochastic Series Expansions (SSE) techniques~\cite{syljuasen_quantum_2002,bauer_alps_2011}, the DTNX $S=1$ Hamiltonian Eq.\eqref{eq:DTNX} is simulated for  $3d$ systems of $N=L\times L/r\times L/r$ sites. For such a weakly coupled chains problem ($J_\perp/J\simeq 0.08$), it is numerically very favorable~\cite{sandvik_multichain_1999} to use anisotropic aspect ratios $r$, depending on the impurity concentration~\footnote{We used $(x,r)=~(\frac{1}{6},6)$, $(\frac{1}{8},8)$, $(\frac{1}{10},10)$, $(\frac{1}{12},6)$, $(\frac{1}{16},8)$, $(\frac{1}{20},10)$ respectively.}. This allows to perform an accurate finite-size scaling analysis using increasing system sizes, with chain lengths varying from $L=24$ up to $L=120$. Disorder averaging is carried out over a large number $\ge 300$ of independent  samples.
\begin{figure}
    \includegraphics[width=.942\columnwidth,clip,angle=0]{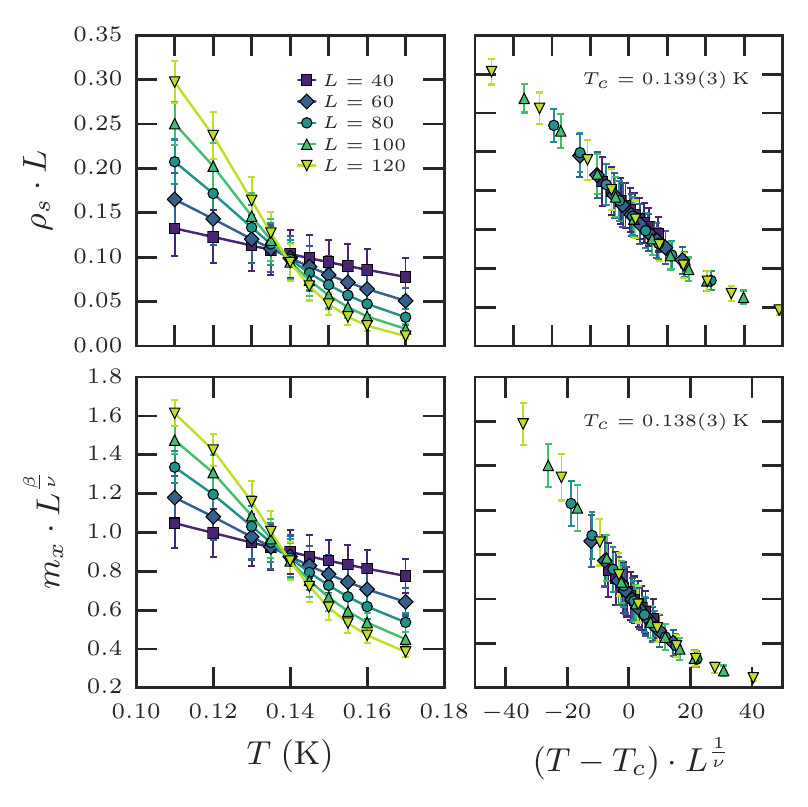}
    \caption{\co Finite size scaling analysis for the disorder average spin stiffness $\rho_s(L)$ (top panels) and  transverse AF order parameter $m_x(L)$ (bottom panels) following the scaling forms given by Eq.~\eqref{eq:stiff_scal_antz}. QMC results obtained for the DTNX Hamiltonian Eq.~\eqref{eq:DTNX} on $L\times L/10\times L/10$ lattices of various sizes at $H=H^*=13.6$~T with $x=10\%$ of impurities.}
     \label{fig:crossings}
\end{figure}

As exemplified in Fig.~\ref{fig:crossings} for $H=H^*$ and $x=10\%$ of impurities, a finite temperature transition  is clearly observed at $T_c=138(4)$ mK using two different estimates: the spin stiffness $\rho_s$~\cite{pollock_path-integral_1987,sandvik_finite-size_1997} and the transverse AF order parameter $m_x=\sum_{i,j}{\rm e}^{i{\boldsymbol{q}}\cdot {\boldsymbol{r}}_{ij}}\langle S^{+}_iS^{-}_{j}\rangle/N^2$ at ${\boldsymbol q}=(\pi,\pi,\pi)$. A standard finite size scaling analysis~\cite{sandvik_computational_2010}
\bea
    \rho_s(L)&=&L^{2-d}\,\, \mathcal{G}_{\rho_s}\left[L^{1/\nu}\left(T-T_c\right)\right]\nonumber\\
    m_x(L)&=&L^{-\beta/\nu}\,\, \mathcal{G}_{m_x}\left[L^{1/\nu}\left(T-T_c\right)\right],
    \label{eq:stiff_scal_antz}
\eea
with $d=3$, and the $3d$-XY critical exponents~\cite{burovski_high-precision_2006,campostrini_theoretical_2006,beach_comment_2005} $\nu=0.6717$ and $\beta=0.3486$, is used to extract $T_c$, after a Bayesian scaling analysis~\cite{harada2011,harada2015}. Both estimates from the stiffness and the order parameter agree very well within error bars~\footnote{Corrections to scaling of the form ${\mathcal{G}}\left[L^{1/\nu}\left(T-T_c\right)(1+cL^{-\omega})\right]$  gives similar values within one standard deviation. Final $T_c$ estimates are averages of the individual $T_c$ from both crossings with and without irrelevant corrections. Error bars reflect uncertainty between various estimates, i.e. min($T_c-\Delta T_c$) and max($T_c+\Delta T_c$).}.

\begin{figure}
    \includegraphics[width=.92\columnwidth,clip,angle=0]{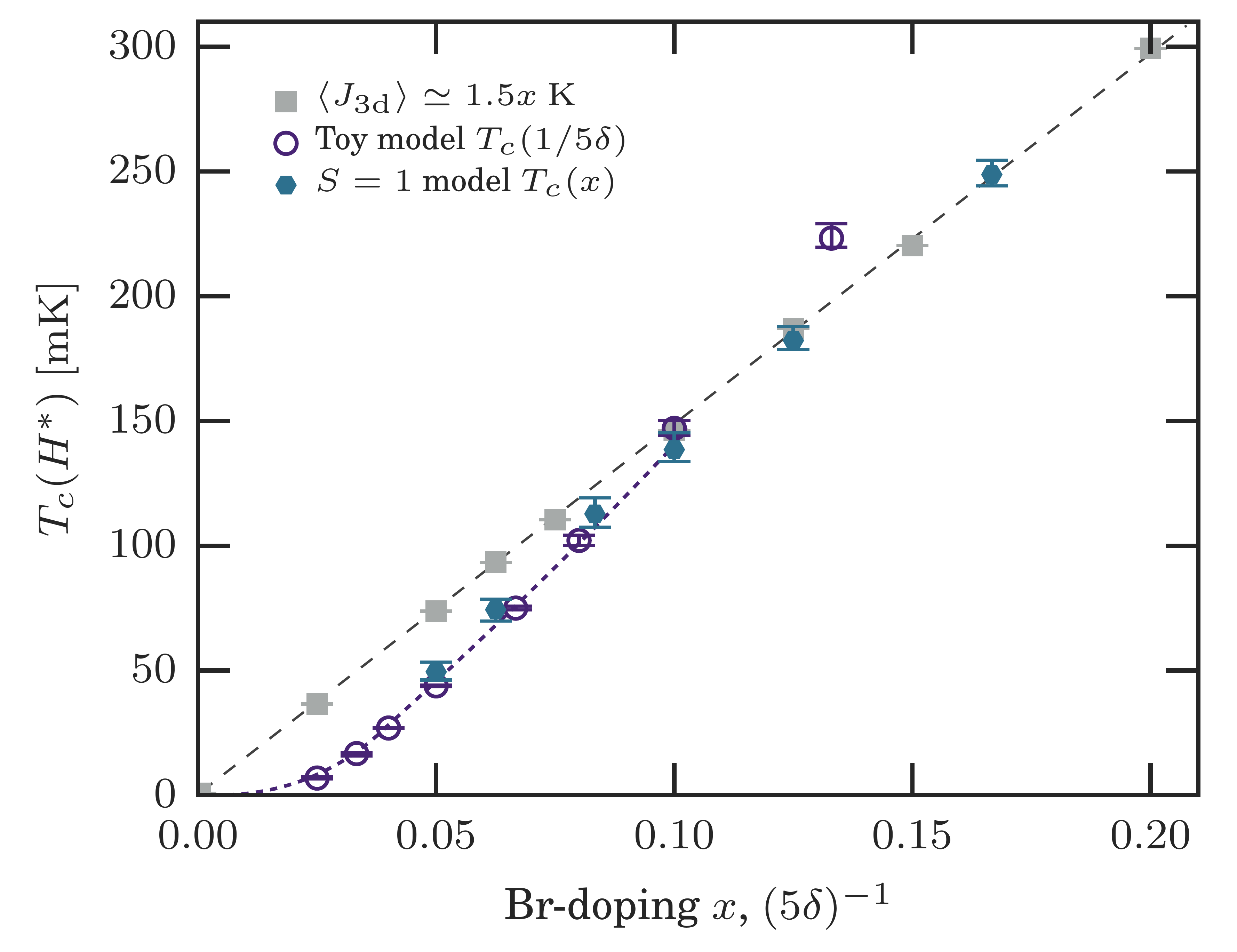}
    \caption{\co Critical ordering temperature for the impurity-induced LRO at $H^*=13.6$ T plotted against the impurity concentration $x$ for the $S=1$ model Eq.~\eqref{eq:DTNX} (hexagon) and for the effective HCB toy-model Eq.~\eqref{eq:eff} at half-filling (circle) plotted against $({5\delta})^{-1}$, suggesting LRO for all finite $x$ values. The average effective pairwise coupling between impurities in the transverse direction is also shown (square) for comparison. Lines are guides to the eyes.}
     \label{fig:tcx}
\end{figure}

Similar simulations and analyses are then repeated for different concentrations $x$ of impurities, still at the crossover field $H^*$, in order to extract the doping dependence $T_c(x,H^*)$. Results are plotted in Fig.~\ref{fig:tcx} for $5\%\le x \le 16.67\%$ where we observe LRO at finite temperature for all doping levels. The ordering temperature grows linearly with $x$. This is qualitatively expected from a naive mean-field reasoning, as the average coupling between the chains (setting the $3d$ energy scale for finite temperature LRO) is $\langle J_{\rm 3d}\rangle\sim J_\perp x$. More precisely, exact diagonalization caculations of the effective pairwise coupling between impurities in DTNX, discussed in Ref.~\cite{orlova_nuclear_2016,dupont_long_2016}, yield an average energy coupling in the transverse direction $\langle J_{\rm 3d}\rangle\simeq 1.5x$~(K), which compares well with QMC estimates, at least for large enough dopings $x\ge 8\%$ (Fig.~\ref{fig:tcx}). For small $x$, accurate estimates for $T_c$ are very hard to obtain because simulations get slower with inverse temperature, and finite size effects become more serious when the number of impurities decreases.
Nevertheless, we can observe at low doping that the ordering temperature starts to deviate form a simple linear scaling and displays a faster decay. While it is impossible to exclude the existence of a critical concentration $x_c< 5\%$ where $T_c$ vanishes,  it is reasonable to expect that $T_c(x,H^*)$ will vanish only when $x\to 0$, presumably with a convex form different from the mean-field-like shape observed for $x> 8\%$.\\

{\it Hard-core bosonic toy-model.---} At this stage, it is  instructive to compare the results obtained for the realistic microscopic DTNX Hamiltonian~\eqref{eq:DTNX} with the simple toy-model HCB Hamiltonian~\eqref{eq:eff} for which QMC simulations have been performed at half-filling ($\mu=0$). Instead of working on a diluted impurity lattice with the exponentially suppressed hoppings derived in Ref.~\cite{orlova_nuclear_2016}, it is easier to investigate the toy-model on a regular cubic lattice made of coupled chains with disordered nearest-neighbor hoppings. In order to mimic the exponentially suppressed effective couplings combined with the random distribution of the distances between impurities in the original $S=1$ problem, we follow Refs.~\cite{sigrist_low-temperature_1996,lavarelo_magnetic_2013} and generate random hoppings from the broad distribution $P(t)\sim t^{-1+{1}/{\delta}}$, with $t\le 2.2~\rm K$ along the chains, and $t\le 0.2~\rm K$ in the transverse directions,  $\delta$ being a phenomenological disorder parameter.

Simulations are carried out for $L\times L/5\times L/5$ lattices with $L=20,\,30,\,40,\,50$, and averaged over a large number $\ge 500$ of samples for $2\le \delta\le 8$. When performing a similar finite-size scaling analysis as explained above, LRO is also detected at low temperature $T_c(\delta)$ which vanishes in the large $\delta$ limit.
More precisely, the disorder parameter $\delta$ of this toy-model can be related to the impurity concentration $x$, such that $1/\delta= 5x$ yields a remarkably good agreement between the two models~\footnote{Following Ref.~\cite{lavarelo_magnetic_2013}, the effective coupling distribution is a power-law with $1/\delta\simeq 2x\xi$ in $d=1$. For $d=3$, our toy-model description gives quantitative agreement with the original $S=1$ model if $1/\delta=2x v_\xi$, with $v_\xi=2.5$ a $3d$ localization volume.}. For this HCB toy-model,  less numerically demanding, one can reach smaller critical temperatures, thus supporting  that $T_c(x,H^*)\to 0$ for $x\to 0$.  This comparison justifies the fact that  impurity-induced LRO at $H^*$ in DTNX is driven by an effective residual interaction, albeit small and random in magnitude, between localized states living on Br-doped bonds. Such a mechanism is analogous to what is generically observed for a wide class of doped spin-gapped compounds~\cite{bobroff_impurity-induced_2009}.

\begin{figure}[t]
    \includegraphics[width=0.92\columnwidth,clip,angle=0]{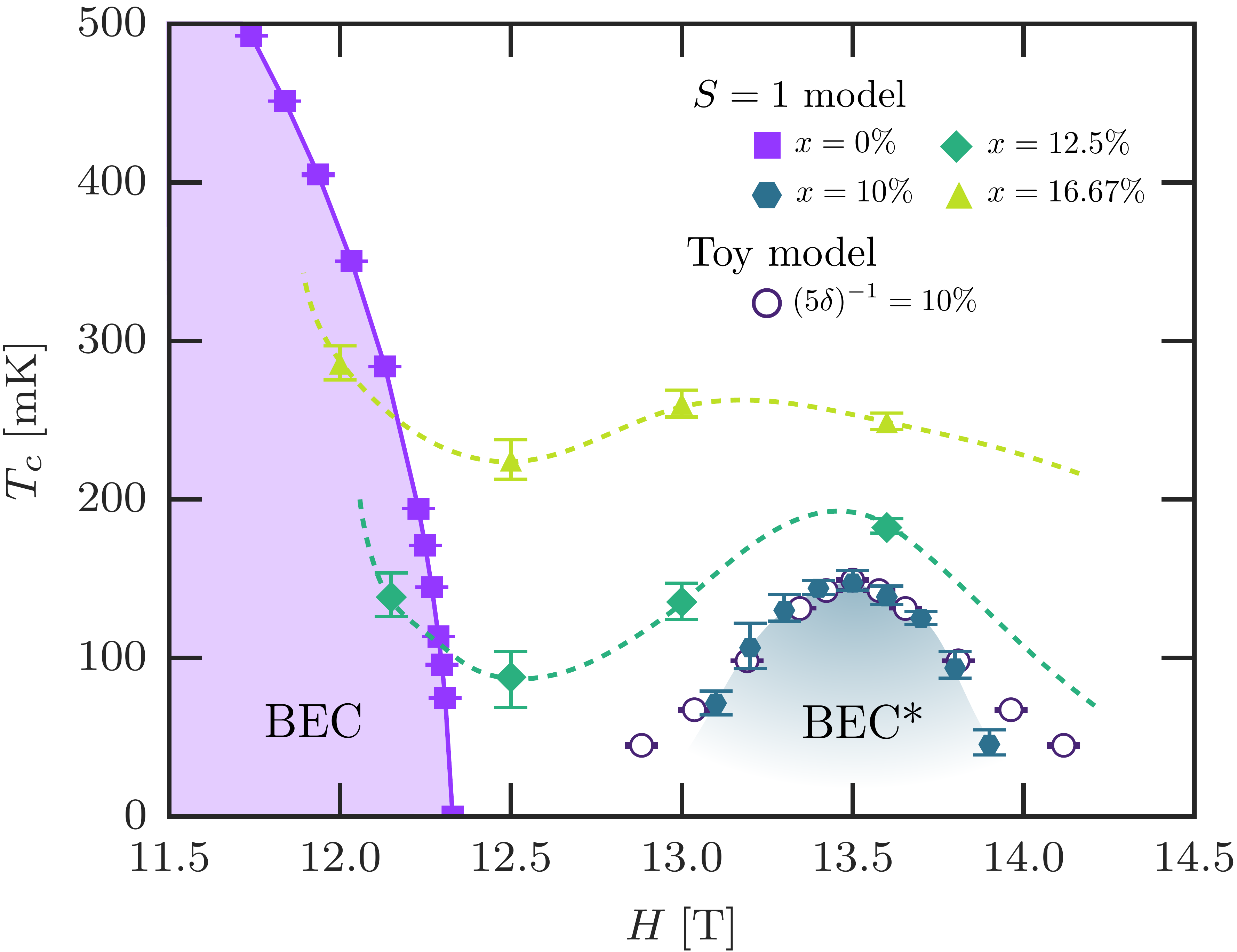}
    \caption{\co~Field $H$ -- temperature $T$ phase diagram for DTNX obtained by QMC simulations for both the $S=1$ model Eq.~\eqref{eq:DTNX} with $x=0$ (square), $x=10\%$ (hexagon), $x=12.5\%$ (diamond), $x=16.67\%$ (triangle) and the HCB effective Hamiltonian Eq.~\eqref{eq:eff} with $\delta=2$ (circle). The HCB BEC$^*$ dome has been centered around $13.5$ T. Dashed lines are guides to the eyes.}
     \label{fig:tch}
\end{figure}

{\it Field-temperature phase diagram.---} The next significant question concerns the possible extension of the ordered regime away from the crossover field $H^*$. Indeed, as schematized in Fig.~\ref{fig:sketch} (d-e), we  expect the level crossing of a single doped dimer to spread and acquire a band-width due to the effective couplings, in analogy with clean weakly coupled dimers~\cite{mila_ladders_1998,giamarchi_coupled_1999}, yielding an extended finite temperature ordered regime around the crossover field $H^*$, dubbed BEC$^*$. In order to address this issue, we have performed QMC simulations of the DTNX model Eq.~\eqref{eq:DTNX} at various values of the external magnetic field $H$ between 12 T and 14 T, for different impurity levels. Results are reported in Fig.~\ref{fig:tch} where the field -- temperature phase diagram is shown. Clearly an extended impurity-induced LRO regime BEC$^*$ is observed, with a maximum slightly shifted below $H^*$~\footnote{This small shift to  $\simeq 13.5$ T is attributed to the effective interactions.}. While for $x=10\%$ this ordered dome seems to reach its left quantum critical boundary at a field value above $H_{c2}$ [also quantitatively supported by the toy-model Eq.~\eqref{eq:eff} at $\delta=2$], leaving room for an intermediate disordered (Bose glass) state~\cite{yu_bose_2012}, this is no longer true at higher doping. Indeed, at $x=12.5\%$  the BEC$^*$ regime overlaps with the low-field BEC dome ($H\le 12.3$ T), excluding the possibility to stabilize an intermediate Bose-glass, and the situation is even more dramatic at $x=16.67\%$.

It is crucial to notice that this effect goes beyond a simple percolation picture. Indeed, the site percolation threshold on a cubic lattice being $p^*\simeq 0.312$~\cite{deng_monte_2005}, one expects an infinite-size Br-doped cluster hosting LRO above a concentration of Br-impurity $2x=p^*$. Therefore, if $x> 15.6\%$ LRO occurs for the entire gapless regime, from low field up to $H'_{c2}=(D'+4J'+8J_\perp)/(g\mu_B)\simeq 16.7$~T. Below this threshold, as for instance seen for $x=12.5\%$ in Fig.~\ref{fig:tch}, the ordering mechanism is controlled by effective couplings beyond nearest-neighbor Br-doped dimers.\\

{\it Summary and discussions.---} An impurity-induced BEC-type AF ordering is expected for DTNX in the vicinity of {$H^*=13.6$~T}, as unveiled by our large-scale QMC simulations performed for the microscopic realistic $S=1$ model Eq.~\eqref{eq:DTNX}. The critical temperature at this crossover field grows with the doping $x$ (Fig.~\ref{fig:tcx}), a result nicely supported by an effective hard-core bosons toy-model description based on localized two-level systems coupled through a random hopping, thus confirming the relevance of the analogy between this disorder-induced BEC$^*$ order and the impurity-induced LRO mechanism observed at zero field for several spin-gapped compounds~\cite{bobroff_impurity-induced_2009}. The temperature and field ranges where this new ordered phase is expected to occur are clearly experimentally accessible for realistic doping levels $x$, either using NMR, neutron scattering, or thermodynamic probes such as specific heat measurement. The experimental observation of this rather exotic disorder-induced BEC$^*$ phase clearly opens new routes to address the interplay between disorder and interactions in such quantum systems.

Numerically, accessing very low $T_c(x,H)$ using QMC simulations, typically below $10$ mK, is very challenging. It is therefore difficult to draw firm conclusions regarding the precise field extension of the BEC$^*$ regime around $H^*$ in the $T\to 0$ limit.  For $x=8\%$, Yu {\it{et al.}}  reported a quantum phase transition into a Bose-glass state above $12.3$ T~\cite{yu_bose_2012}.
This is in agreement with our estimate for the onset of the overlap between BEC and BEC$^*$ domes, expected to be experimentally detectable for $x> 10\%$. However we stress here that this reported Bose-glass state at $x=0.08$ may only exist in a very narrow field regime between the BEC and BEC$^*$ ordered states. At lower impurity concentration levels, we further expect a more extended and experimentally accessible BG regime intervening between two ordered phases.\\

Upon increasing further the field $H>H^*$, we then expect the BEC$^*$ dome to eventually vanish, presumably before $H'_{c2}\simeq 16.7$~T for $x< 15.6\%$ (where the Br-doped cluster reaches its percolation threshold), thus offering the possibility to stabilize another Bose-glass state at high magnetic field, before the complete saturation of the spins. It is therefore quite promising to contemplate this BEC -- Bose glass criticality at such a high-field transition where no surrounding order would spoil its genuine  properties, allowing to determine its critical exponents which are still controversial~\cite{yu_quantum_2012,Yao2014,yu2014comment}. For other experimental systems with a magnetic Bose glass regime (for a recent review, see Ref.~\onlinecite{zheludev_dirty-boson_2013}) one could expect a similar disorder-induced BEC revival~\cite{nohadani_bose-glass_2005} provided the separation of energy scale between clean and doped sites is large enough, as for instance in the metal-organic spin ladder (Hpip)$_2$CuBr$_{4(1-x)}$Cl$_{4x}$~\cite{ward_spin_2013}.

\begin{acknowledgments}
We would like to thank M Horvati\'c and the NMR team from LNCMI-Grenoble for highly valuable experimental inputs, as well as T. Roscilde for discussions.
This work was performed using HPC resources from GENCI (Grant No. x2015050225 and No. x2016050225), and is supported by the French ANR program BOLODISS and R\'egion Midi-Pyr\'en\'ees.
\end{acknowledgments}

%

\end{document}